# Color2Struct: efficient and accurate deep-learning inverse design of structural color with controllable inference

Sichao Shan[1], Han Ye[2], Zhengmei Yang[1†,] Junpeng Hou[3#], Zhitong Li[1*]

[1]State Key Laboratory of Information Photonics and Optical Communications, School of Science, Beijing University of Posts and Telecommunications, Beijing 100876, China

[2]School of electronic engineering, Beijing University of Posts and Telecommunications, Beijing 100876, China

[3]Pinterest Inc., San Francisco, California 94107, USA

†yangzm@bupt.edu.cn, # jhou@pinterest.com, *zhitong.li@bupt.edu.cn

## Abstract

Deep learning (DL) has revolutionized many fields such as materials design and protein folding. Recent studies have demonstrated the advantages of DL in the inverse design of structural colors, by effectively learning the complex nonlinear relations between structure parameters and optical responses, as dictated by the physical laws of light. While several models, such as tandem neural networks and generative adversarial networks, have been proposed, these methods can be biased and are difficult to scale up to complex structures. Moreover, the difficulty in incorporating physical constraints at the inference time hinders the controllability of the model-predicted spectra. In this work, we propose Color2Struct, a universal framework for efficient and accurate inverse design of structural colors with controllable predictions. By utilizing sampling bias correction, adaptive loss weighting, and physics-guided inference, Color2Struct improves the prediction of tandem networks by 65% (color difference) and 48% (short-wave near-infrared reflectivity) in designing RGB primary colors. These improvements make Color2Struct highly promising for applications in high-end display technologies and solar thermal energy harvesting. In experiments, the nanostructure samples are fabricated using a standard thin-film deposition method and their reflectance spectra are measured to validate the designs. Our work provides an efficient and highly optimized method for controllable inverse design, benefiting future explorations of more intricate structures. The proposed framework can be further generalized to a wide range of fields beyond nanophotonics.

## Introduction

Machine learning (ML), especially deep learning (DL), has revolutionized many fields in science, including materials design[1–6], drug discovery[7–13], and protein folding[14–22]. In materials science, ML and DL accelerate the discovery of new materials and optimize their properties. In drug discovery, DL models analyze vast biomedical datasets to identify potential drug targets and molecular structures, thereby shortening development cycles. In protein folding, breakthroughs in DL technology, such as AlphaFold[17,20–22], have achieved unprecedented accuracy in protein structure prediction. Similarly, in the fields of optics and photonics, DL is fostering numerous innovations. For example, optical neural networks (ONN) combine optical signal processing with DL to enable fast and efficient data processing[23–29]. Additionally, DL-based inverse design methods optimize the structure of nanophotonic devices, allowing precise control over optical properties. More recently, it has been shown that DL models can help automate the design of structural colors[30–37]. Structural color arises from the scattering and interference of light within nanostructures[38–42]. Since colors are governed by nanostructures and the material properties rather than chemical dyes, structural colors offer greater durability, tunability, and stability under varying environmental conditions. For the multilayer structures, when incident light propagates in the layers, electromagnetic waves at different wavelengths undergo constructive or destructive interference at the interfaces, which can selectively enhance or diminish specific wavelengths, thereby producing highly saturated colors[43–46]. Compared to metasurfaces or photonic crystals that require time-consuming fabrication procedure, multilayer structures are more fabrication friendly. Consequently, it hold significant promise for applications in high-quality display technologies[40,44,47–50], color printing[43,44,51–55], and optical encryption[56–58]. The design of multilayer structural colors is highly complex, involving comprehensive trade-offs among multiple degrees of freedom, such as the arrangement of the layers, and the thickness and materials of each layer. Traditional design methods rely on extensive parameter tuning and repeated simulation validation, making the process not only time-consuming but also heavily dependent on the designer's experience and intuition. By utilizing their efficient algorithms and big data processing capabilities, DL models can effectively learn the complex nonlinear relations between nanostructure parameters and the optical properties of the materials[45,46]. Different methods such as tandem networks and generative adversarial networks have been adapted to design various classes of multilayer structures with different materials[59–67]. These DL-based approaches greatly reduce the reliance on traditional physical simulations and the designer's experience, while significantly improving design efficiency.

However, many important modern DL techniques related to data sampling, model training and inference optimization remain underexplored, even though they have been the key factors in driving the success of DL over the past decades, as exemplified by recent breakthroughs in large language models (LLMs)[68–74]. This could hinder the future applications of DL algorithms for solving larger-scale real-world designs. To start with, one issue arises from the severe biases in the model. For example, while the model may make accurate predictions for some specific colors, it performs poorly when predicting colors with high purity and saturation. Furthermore, simply scaling up the models is not always the most efficient way to improve performance or tackle more complex designs. For instance, with an increasing number of layers in the structure, deeper and wider networks do not guarantee better predictions. Last but not least, current studies often focus solely on whether the predicted nanostructure parameters can achieve a desired point in the chromaticity space[75–78], overlooking the characteristics of the optical spectra and other physical properties or design constraints. For example, reducing reflectivity in the short-wave near-infrared range can enhance light absorption, which is particularly beneficial for solar photovoltaic absorber materials. Following these leads, we arrive at three fundamental questions. 1) Are there any biases in the current DL models for inverse design of structural colors? If so, what's the origin of the bias and how it to be mitigated? 2) When the structure becomes more intricate with gigantic parameter spaces, how should we improve model performance, given that simple scaling-up is insufficient? 3) How can we efficiently incorporate the physical properties and constraints of the spectra during inference time to meet requirements for various applications?

In this work, we comprehensively address the above questions by closely examining all stages, including training data generation, model training, and inference, and by introducing Color2Struct, a framework for efficient, accurate, and controllable inverse design of structural colors. By investigating a specific design problem involving a five-layer nanophotonic structure, we reveal severe biases in a tandem network and demonstrate that merely scaling up the model is not efficient for solving complex problems. We identify the primary sources of bias as uneven distribution of the training data in the color space and inadequate training strategies. To mitigate the bias, we introduce Sampling Bias Correction (SBC) to balance data distribution and Adaptive Loss Weighting (ALW) to better guide the learning process. By combining the two methods, we improve model performance on the test dataset, achieving up to a 57% reduction in average color difference. For the third question, we adopt the principle of inference time scaling to allow multiple prediction attempts, and most importantly, we propose Physics-Guided Inference (PGI) to embed spectral information into input features and achieve precise control over the predicted results[79]. We demonstrate that these methods significantly improve the prediction performance for high-purity colors, including the primary RGB colors, achieving a 65% improvement compared to existing studies, and notably reduce the reflectance in the short-wave near-infrared range, with a maximum reduction of 48%. We also apply Color2Struct to different design patterns with varying numbers of layers and different candidate materials and observe similar gains in prediction quality. Finally, we fabricate alternating $SiO_2$/Ti multilayer structures using plasma-enhanced chemical vapor deposition (PECVD) and e-beam evaporation[80,81]. The normal-incidence reflection spectra of the samples are measured under ambient conditions using a UV–Vis–NIR spectrophotometer to validate the proposed design[82]. The experimental results exhibit excellent agreement with the model predictions, which strongly validates the feasibility and effectiveness of the proposed Color2Struct framework.

## Results

### The problem setup

Before presenting our optimized framework, we start with a multilayer structure as shown in Fig. 1(a) to introduce the setup of the inverse design problem. From top to bottom, this structure consists of five alternating layers of $SiO_2$ and Ti, stacked on a Si substrate, with varying thicknesses. It forms two embedded Fabry-Perot (FP) cavities due to the refractive-index contrast at the interfaces. A fixed-thickness aluminum (Al) layer at the bottom acts as a high-reflectivity back reflector. The arrows indicate the light reflected from the surface, and we focus on the color derived from the reflection spectrum. A cross-sectional Scanning Electron Microscopy (SEM) image of the experimental sample is shown on the right[83]. Note that the top Al layer in the SEM image is not part of the multilayer structure but a protective layer during imaging. The detailed imaging method can be found in the Methods section.

The exact data workflows are illustrated in Fig. 1(b), and in this work, we set the maximum thickness of metals and dielectrics as 50 nm and 500 nm, respectively. In current studies, datasets are typically prepared by calculating the spectra of these structures using the Transfer Matrix Method (TMM)[84–86]. These spectra are then converted into color coordinates, such as the CIE-1931-XYZ (CIE-XYZ) or CIE-1976-Lab (CIE-Lab) color spaces, using color matching functions (CMFs)[75–78,87–89]. Each record in the dataset consists of the nanostructure parameters and the corresponding color coordinates, which are also denoted as the features and labels in the context of supervised learning.

For demonstration purposes, we use the tandem network, which consists of two parts, the forward neural network (FNN) and the inverse neural network (INN). The FNN, shown in Fig. 1(c), takes the structural parameter $D$ as input and outputs the CIE-Lab parameters. As shown in Fig. 1(d), the INN takes the CIE-Lab parameters as input and outputs the structural parameter $D$, which is then passed to a pre-trained FNN. During the training of the INN, the parameters of the FNN remain fixed. Therefore, the model is trained using two steps and only the INN is used to make final design predictions.

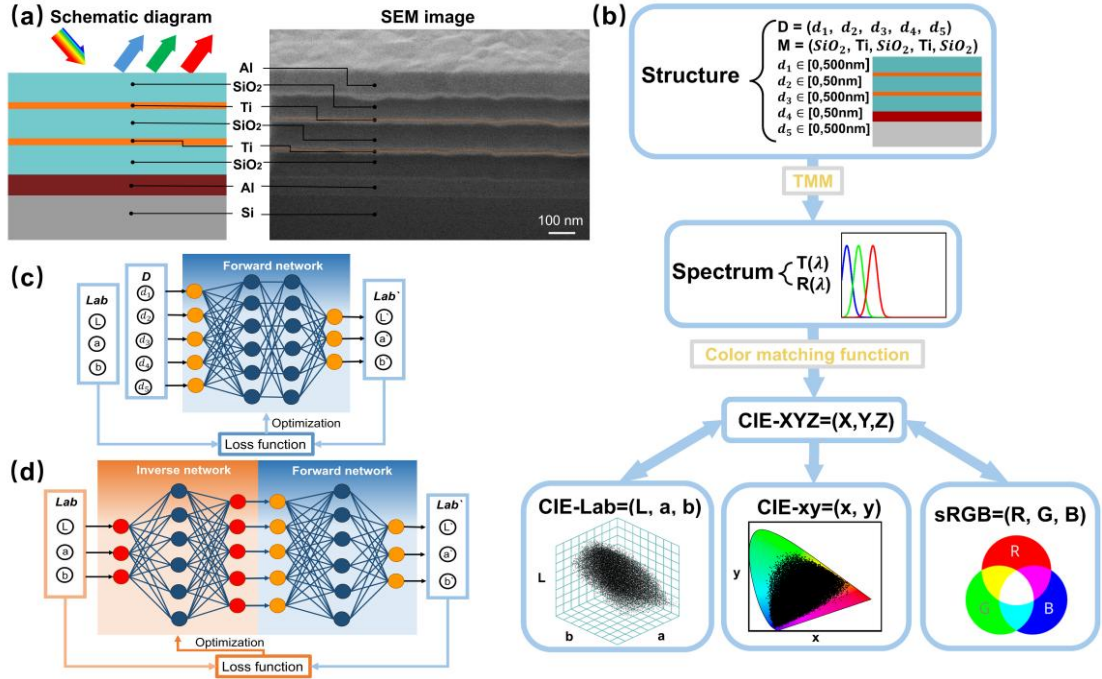

**Fig. 1** (a) Schematic diagram and SEM image of the multilayer structure with alternating $SiO_2$ and Ti layers. The sample's five layers measure 133 nm, 8 nm, 136 nm, 18 nm, and 148 nm, respectively. All thickness is obtained via model predictions. The top and bottom Al layers are 120 nm and 100 nm thick, respectively. The Al layer at the top of the SEM image is not part of the multilayer structure, but a protective layer during imaging. (b) Diagram of the data workflow. The multilayer structure can be parameterized by a vector $D$ of five real numbers, representing the thickness of each layer. Given a specific $D$, the reflection spectrum $R$ or transmission spectrum $T$, which are functions of wavelength $\lambda$, can be computed using TMM. Then the CIE-XYZ coordinates can be calculated using CMFs and can be further transformed into other color spaces, including CIE-Lab, CIE-xy, and sRGB. The bidirectional arrow indicates that the transformation is reversible. (c) Architecture of the FNN, which consists of an input layer of $D$, hidden layers, and output layers to predict the CIE-Lab. (d) Architecture of the bidirectional network with an INN connected to the pretrained FNN, with its parameters frozen during the training of the INN.

## The challenges

After generating a dataset of 500k records, we train multiple tandem networks with different numbers of layers and varying numbers of neurons per layer. The results are summarized in Fig. 2. Detailed training methods can be found in the Training Details section of the Supplementary. We use color difference as the metric to evaluate the prediction results, as it measures the Euclidean distance between prediction and target in the CIE-Lab coordinates and usually, $\Delta E < 1$ is indistinguishable to the human eye[90,91]. The detailed $\Delta E$ standards can be found in the Metric section of Supplementary. As plotted in Fig. 2(a) and (b), we observe that with deeper and wider FNNs, $\Delta E_{avg}$ (average of $\Delta E$ in test datasets) can be optimized to be less than 1, with the majority of results shift to the orange bar. We also use a pretrained FNN to train several INNs and present the results in Fig. 2(c) and (d). The results indicate that the INNs can make reasonable inferences with $\Delta E_{avg}$ smaller than 1. Nevertheless, when we apply the trained models to make predictions of certain colors with high purity, the performance degrades significantly and in some of the worst cases $\Delta E$ exceeds 10, which is typically deemed unacceptable in structural color design. The model's prediction results for several high-purity red, green, and blue colors are shown in Table S1 in the Supplementary. Such an observation provides direct evidence of inherent biases in the existing methods.

Before exploring the root cause of such a bias, we first notice that $\Delta E_{avg}$ alone is inadequate to capture the model performance. Thus, we also track $\Delta E_{max}$ in Fig. 2. It is important to note that the original objective is to describe the maximum $\Delta E$ in the model's predictions. However, to reduce the impact of extreme predictions on model evaluation, the dataset is divided into small batches. The maximum $\Delta E$ within each batch is then averaged, and this average is defined as the $\Delta E_{max}$. For the FNNs, both $\Delta E_{max}$ and $\Delta E_{avg}$ exhibit similar descending trends as the model become more complex, yet the former is much worse. The gap between the two metrics in the INNs is even more pronounced, with $\Delta E_{max}$ deteriorating noticeably as the number of layers/neurons exceeds 6/1000. This can be ascribed to the fact that INNs are generally limited by the pretrained FNNs, which also suggests that the bias likely originates from the training data. This indicates that the model becomes "overfitted" on the easier examples, and the training data is inherently imbalanced in terms of example difficulty.

Additionally, the spectral characteristics in predictions are associated with a specific physical quantity. In this work, we will use the reflectance integral over a portion of the near-infrared range (1000 nm – 1200 nm), denoted as the new metric $R$, to assess the convergence of the reflectance spectra in the short-wave near-infrared range. In our tests, the performance of $R$ is very low and current research typically only focuses on chromatic information. We add $R$ as an additional input during the training process. The modified model architecture and the corresponding test results are shown in Fig. S1 and Table S2 of the Supplementary, respectively. While the performance of $R$ improves in the test results, there is still room for further optimization. Further analysis suggests that the issue primarily arises because the input features used during inference lack spectral characteristic information.

To systematically address the above challenges, we will present Color2Struct and show how it drastically improves the model performance and provides precise control over the prediction.

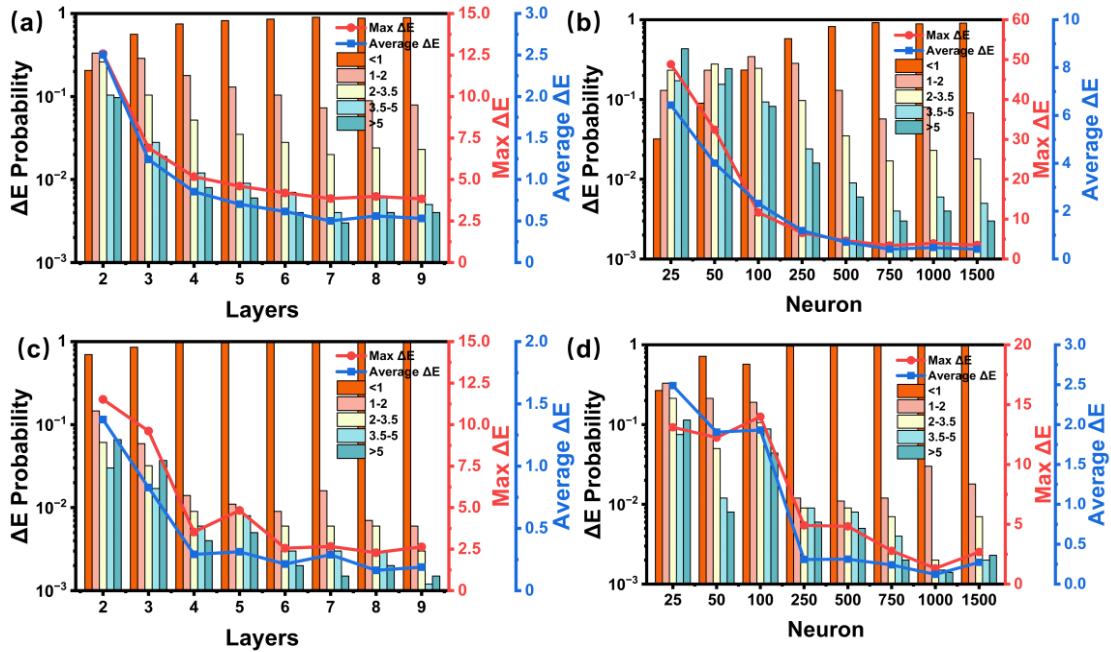

**Fig. 2** Testing metrics for varying model architectures. (a) The histograms show the probability distribution of $\Delta E$ with respect to the number of layers in the FNN. The blue and red curves show the $\Delta E_{avg}$ and $\Delta E_{max}$. (b) Similar to (a) but plotted with respect to the increasing number of neurons in a single layer. From the panels, it is clear that the benefits of scaling up vanish when the number of layers/neurons becomes larger than 7/500. (c) and (d) Similar to plots (a) and (b) but for INN training. The same trends for scaling up are observed, but the performance gains plateau when the number of layers/neurons exceeds 4/250.

### Color2Struct

In Color2Struct, we carefully consider the design requirements in the context of nanophotonics and aim to provide a toolbox for improving training efficiency and inference performance in a systematic way that supports diverse models in DL-based optical design and related tasks. We will start with the data generation, then the training strategy and finally the inference, and show how it can be applied to the tandem network.

### Sampling Bias Correction

Currently, it is a common practice to construct datasets by uniformly sampling in the nanostructure parameter space, since direct sampling in the color space is infeasible. Due to the complex nonlinear relations between nanostructure parameters and the color space coordinate (CIE-Lab), the obtained data exhibits a highly uneven distribution in the color space, which is demonstrated in Fig. 3(a) and (b). The mesh in Fig. 3(a) clearly shows the concentration of data (orange parts), which becomes more skewed in the CIE-Lab color space.

Most importantly, as illustrated by the fitted data distribution (orange curve) in the CIE-xy space of Fig. 3(b), the data is primarily concentrated in low-purity color regions and becomes increasingly sparse in high-purity regions (e.g., the rightmost corner of the sRGB triangle). This is also consistent with the poor model performance on high-purity colors as discussed in the previous section. Such a skewed distribution negatively impacts the model's learning, causing it to overly focus on the more frequent examples and ultimately translating into bias in the models.

In Color2Struct, we apply the SBC to reduce the inherent biases in the dataset. More specifically, we resample the data in the CIE-Lab space by dividing the color space into many grids with finer granularity, and then we randomly retain one data point within each grid. This straightforward treatment does not assume any prior knowledge of the mapping between the two spaces and is shown to be very effective in producing a more uniform distribution of the CIE-Lab data. In Fig. 3(c) and (d), we plot the data distribution after SBC with grid size $1 \times 1 \times 1$. While it is apparent in the plot, we provide a more quantitative characterization by comparing the standard deviation ($\sigma$) of the Gaussian fitting curves in Fig. 3(b) and (d). The $\sigma$ values along the x- and y-axes increase by 139% and 142%, respectively, from 0.0469 and 0.0563 to 0.1123 and 0.1367, indicating that the data distribution becomes more uniform after applying SBC.

The benchmarking results of models trained using the new dataset are summarized in Fig. 4, and we observe that, for the FNN $\Delta E_{avg}$ and $\Delta E_{max}$ are reduced by 33% and 48%, respectively, on the test dataset. More critically, we also observe better performance in the INN, with the two metrics improving by 53% and 62% on the test dataset. The results shown in the main text are the test results of each model on the test dataset. The detailed results and loss curves on the training and validation datasets for different optimization methods are presented in the Supplementary Fig. S2 and Fig. S3, respectively.

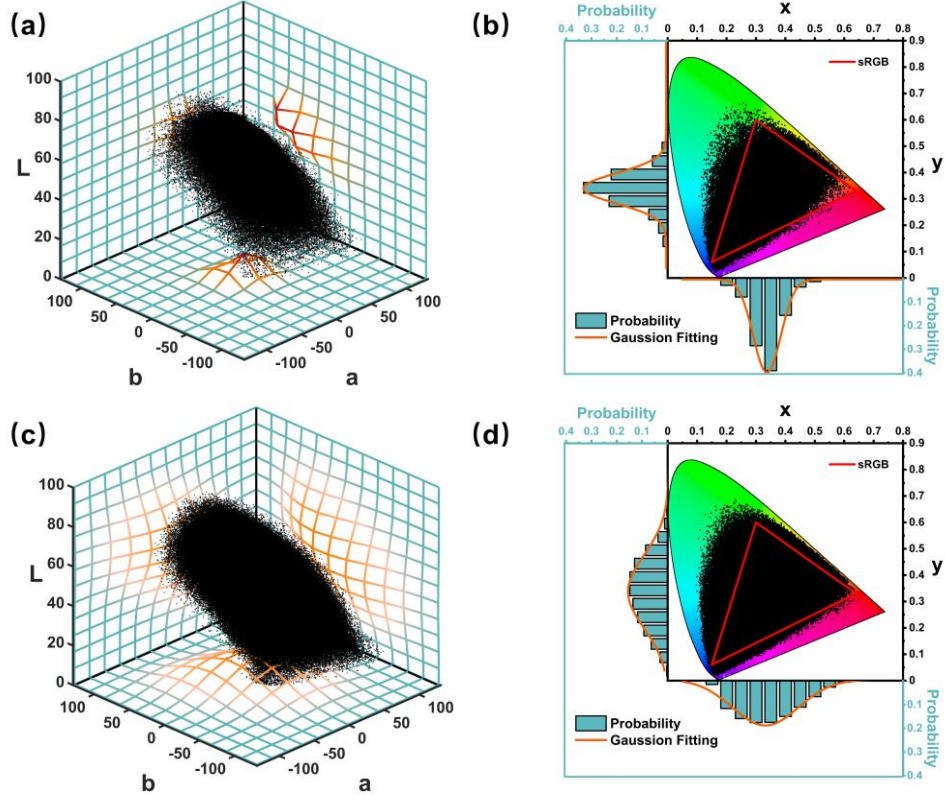

**Fig. 3** (a) Data distribution in CIE-Lab obtained from original training dataset. The meshes represent the projection of the 3D data points onto 2D planes in the color space. (b) Plot of the same data as in (a), shown in the CIE-xy chromaticity diagram. The orange curves illustrate the projection of the data along the x- and y-axes, while the red triangle highlights the boundary of the sRGB color space. Colors positioned closer to any of the three corners have higher purity. (c) and (d) Similar to (a) and (b) but plotted for data after SBC. It is clear that the dataset exhibits more uniform distribution with SBC. By comparing panels (b) and (d), the standard deviation of the Gaussian fitting along the x- and y-axes increases by 139%/142%, respectively.

**Adaptive Loss Weighting**

During the training, we add $R$ as an additional input and use the standard mean squared error (MSE) $L_{MSE} = \frac{1}{B}\sum_i MSE(LabR, LabR`)$, as the loss function. Here $LabR$ is the ground truth value in the test dataset, $LabR`$ is the output of the FNN, and $B$ is the size of a mini-batch. The model tends to focus on learning the majority and simpler examples in each batch. However, correcting the bias in the training data alone is not sufficient, as discussed in the previous paragraph. Inspired by hard-negative mining and focal loss[92–94], we introduce ALW method in Color2Struct. Instead of assigning the same weights to all examples in a mini-batch, we dynamically change the weights based on the model's status, allowing the model to focus more on the harder examples. More specifically, we use $\Delta E_i$ and $R$ to compute the weighted MSE loss. So, the new loss function becomes:

$$L_{ALW-MSE} = \frac{1}{B}\sum_i (1+\alpha)(1+\beta)MSE(LabR, LabR`)$$

$$\alpha = \gamma + \frac{\Delta E_i}{\Delta E_{batch\_max}}, \quad \beta = \frac{1}{R_{i\_truth}}$$

where $\Delta E_i$ and $R_{i\_truth}$ denote the color difference and the ground truth value of $R$ for each sample in a mini-batch, $\Delta E_{batch\_max}$ is the maximum color difference within the batch, $\gamma$ is a hyperparameter used to avoid vanishing dynamic weights and training instability. We fine-tune $\gamma$ and find that the optimal value is 0.25 for FNN and 0.5 for INN (see Fig. S4 in Supplementary for details). The corresponding model performance is also summarized in Fig. 4. By applying ALW alone to FNN, we observe reductions of 4% in $\Delta E_{avg}$ and $\Delta E_{max}$ on the test dataset. For the INN, the two metrics are reduced by 8% and 4%, respectively.

Taking a step further, we combine the two treatments in Color2Struct and observe significant performance improvements, indicating a strong synergy between the two treatments. Also in Fig. 4, the FNN trained with both SBC and ALW achieves a reduction of 46% and 54% in $\Delta E_{avg}$ and $\Delta E_{max}$ on the test dataset, while the INN achieves reductions of 57% and 63%. These improvements indicate that the new loss function can effectively guide the model's learning process, enabling it to learn from a less biased dataset.

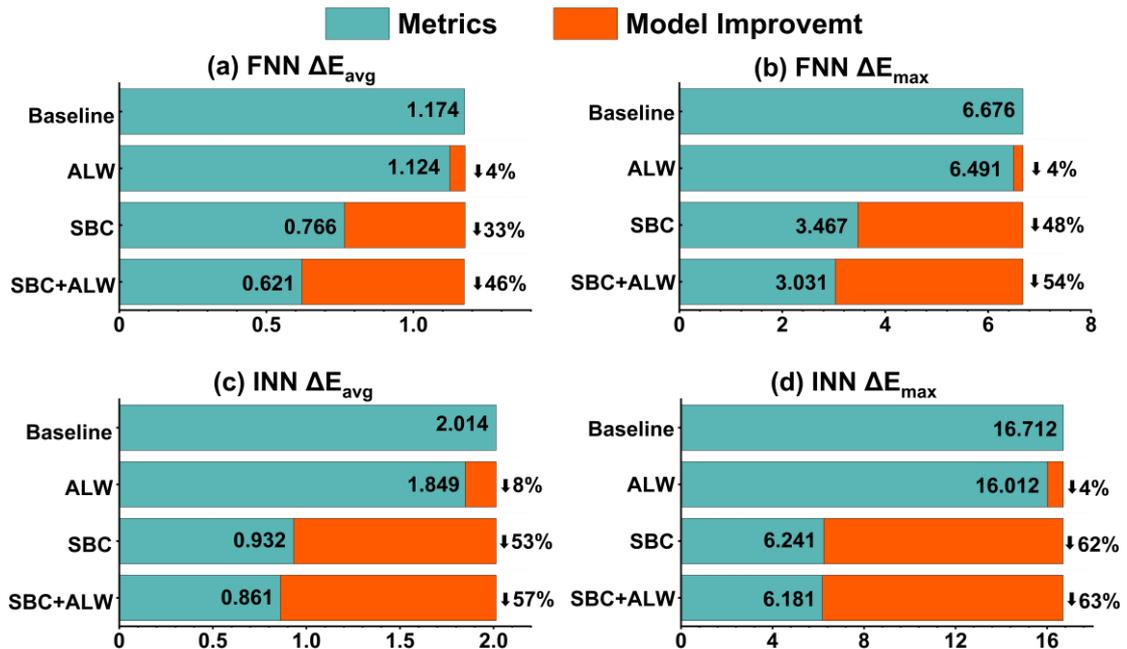

**Fig. 4**(a) and (b) present benchmark performance of the FNN model on the test dataset, with SBC and ALW applied in Color2Struct. Cyan bars represent the metric values, while orange bars show performance improvements compared to the baseline. The synergy between SBC and ALW leads to significantly better results than their separate treatments. (c) and (d) Similar to (a) and (b), but for the INN.

**Physics-Guided Inference**

As we now have satisfactory performance in terms of color difference, we turn to consider the physical and application constraints of the design. As discussed in previous subsections, we focus on the new metric $R$, which is the integral of reflectivity over a selected portion of the short-wavelength near-infrared range. The baseline results for predicting primary RGB colors are presented in Fig. 5(a). Although the method of adding $R$ as an additional input has partially reduced $R$, the baseline results show reasonable performance in $\Delta E$. However, $R$ remains very high. Inspired by the scaling of inference-time computation in LLMs, we propose and apply PGI to encode such physical constraints into the input features implicitly and scale inference, further achieving performance improvements.

In Color2Struct, PGI comprises two parts. The first part involves performing multiple inferences using Proximity Space Sampling (PSS), which generates proximate input features given the targets. A naive method is to perform a random sampling around the target coordinates. This simple method tends to be quite effective as shown in Fig. 5(a). With 200 random proxies, for example, for the color red, the $\Delta E$ decreases by 5% and $R$ is reduced by 4%; for green, the $\Delta E$ decreases by 9% and $R$ is reduced by 18%. Although the $\Delta E$ for blue increases by 10%, $R$ is significantly reduced by 37%. However, such a method quickly reaches a bottleneck, as we only observe marginal gains with additional samples, and the results become less relevant when the sampling region is expanded.

To generate the proxy inputs more efficiently and to incorporate physical constraints, the second part of PGI employs mathematical line shape functions to simulate reflection spectra and convert them to the CIE-Lab format using CMFs. These mathematical line shape functions effectively control the peak positions and full width at half maximum (FWHM), while also controlling reflectance in the short-wave near-infrared region. Additionally, since the peak position of the reflection spectrum largely determines the chromaticity of the color, the input features generated through this method not only meet the requirements for target color coordinates but also carry the desired spectral characteristics. In particular, we use Standard Line Sampling (SLS) and Generalized Line Sampling (GLS), which utilize standard line shape functions (e.g., Gaussian) and generalized line shape functions (e.g., asymmetric-Gaussian) to simulate reflection spectra. From the benchmark results shown in

Fig. 5(a), both SLS and GLS achieve significant improvements across all metrics for all colors. For red, the $\Delta E$ and $R$ are reduced by up to 13% and 64%, respectively; for green, by up to 63% and 48%, respectively; and for blue, by up to 19% and 53%. Most importantly, we did not require tradeoffs between color distance and long-wavelength reflections as both can be optimized via PGI. Another interesting observation is that while SLS performs better for green and blue, GLS shows the best results for red. This might result from the physical limits of the given structure and suggests that there could be further opportunities by scaling the computation time and mixing different line shape functions.

To visually demonstrate the improvements of Color2Struct in reducing color differences and lowering short-wave infrared reflectance, we plotted the full reflectance spectra and chromaticity diagrams for three primary colors. Here, "Best Result" represents the optimal outcomes from SLS and GLS with low $R$ values. Fig. 5(b) shows the predicted reflectance spectra for a red target color, using the baseline, PSS, and Best Result methods; the corresponding CIE-xy coordinates are provided in Supplementary Fig. S5. Since the peak wavelength and peak reflectance of a spectrum largely determine the perceived color, the peak wavelengths of the three spectra occur at 704 nm, 696 nm, and 688 nm, with reflectance magnitudes of 0.86, 0.86, and 0.87, respectively, and they almost completely overlap. As a result, all three appear red, but the Best Result method achieves the highest purity, while the baseline exhibits the lowest. To more clearly illustrate the optimization effect of PGI on $R$ value, the right-hand panel of Fig. 5(b) magnifies the 1000 nm – 1200 nm region. In this range, the $R$ values for the predictions of the baseline, PSS, and Best Result methods are 8.731, 8.339, and 3.218, respectively, with all three methods yielding reflectance magnitudes below 0.06, and the reflectance result after PGI optimization dropping below 0.02. Fig. 5(c) and (d) repeat this analysis for green and blue target colors. For green, the peak wavelengths are 544 nm, 546 nm, and 541 nm, with peak reflectance magnitudes of 0.88, 0.89, and 0.87, and corresponding $R$ values of 7.563, 6.399, and 3.430. For blue, the peak wavelengths occur at 464 nm, 467 nm, and 466 nm, with peak reflectance magnitudes of 0.83, 0.79, and 0.76, and $R$ values of 3.909, 2.977, and 1.275. The green and blue results follow the same trend observed for red. It is evident that the reflection spectra predicted by the model, using features processed with PGI, exhibit extremely low reflectance in the short-wave near-infrared range, while also demonstrating outstanding performance in predicting high-purity colors.

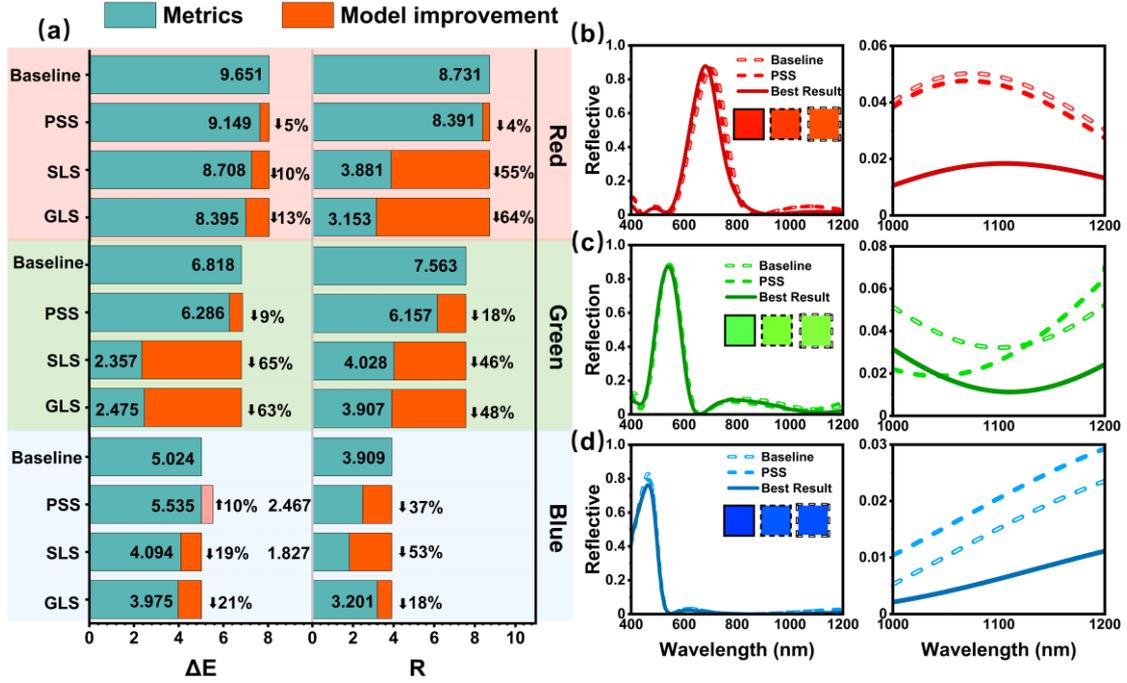

**Fig. 5**(a) shows the benchmark results for $\Delta E$ and $R$ of primary RGB colors using Color2Struct with PGI. The $\Delta E$ and $R$ here are averaged over multiple test datasets. Similar to Fig. 4, cyan and orange bars represent the metric values and performance improvements relative to the baseline. The Baseline uses the target color as the input. PSS performs multiple inferences using random samples around the target color coordinates in CIE-Lab. SLS and GLS adopt standard and generalized line shape functions to generate the input features. Clearly, PGI can better control the physical constraints and reduce color difference in parallel. (b) The left side shows the near-infrared reflectance spectra corresponding to the optimal results obtained using different optimization methods for the target color red, with the inset displaying the corresponding chromaticity diagram. The hollow-dashed, solid-dashed, and solid lines represent the Baseline, PSS, and Best Results, respectively. On the right, a zoomed-in plot of the reflectance spectra in the range of 1000 nm – 1200 nm is provided. (c) and (d) show the same plots, but for the target colors green and blue, respectively.

To demonstrate that our framework can optimize inverse design tasks for various multilayer structures, we modify the number of layers, and the metal and dielectric materials based on the existing structures in our study. The detailed nanostructure schematic diagrams are shown in Supplementary Fig. S6. We then summarize the performance improvements achieved by using this framework for inverse design tasks across three new multilayer structures. Table 1 shows the reductions in $\Delta E_{avg}$ and $\Delta E_{max}$ during training time, as well as the performance improvements in $\Delta E$ and $R$ when PGI is applied to process the input features at the inference time. These results demonstrate that our framework optimizes inverse design tasks for different multilayer structures, with detailed data available in Supplementary Fig. S6–S15.

Table 1. Performance Improvements of Color2Struct in Various Inverse Design Tasks

| Struct | 7Layers | | SiO$_2$-Cr | | Si$_3$N$_4$-Ti | |
|---|---|---|---|---|---|---|
| Metrics | $\Delta E_{avg}$ | $\Delta E_{max}$ | $\Delta E_{avg}$ | $\Delta E_{max}$ | $\Delta E_{avg}$ | $\Delta E_{max}$ |
| FNN | 42% | 47% | 40% | 46% | 38% | 37% |
| INN | 77% | 71% | 41% | 43% | 62% | 57% |
| Metrics | $\Delta E_{avg}$ | $R_{avg}$ | $\Delta E_{avg}$ | $R_{avg}$ | $\Delta E_{avg}$ | $R_{avg}$ |
| Blue | 54% | 58% | 52% | 57% | 62% | 53% |
| Green | 45% | 37% | 16% | 44% | 28% | 63% |
| Red | 14% | 58% | 46% | 51% | 6% | 11% |

**Fabrication and measurement**

In the experiment, we fabricate the blue and red samples using thin-film deposition techniques (see Methods for details). The layer thicknesses of each sample are obtained from the predictions of the trained model. For the blue sample, the layer thicknesses are 133 nm, 8 nm, 136 nm, 18 nm, and 148 nm, while for the red sample, they are 232 nm, 7 nm, 229 nm, 15 nm, and 216 nm. The reflectance spectra of the experimental samples are measured using a reflectance measurement setup. The predicted reflectance spectra are calculated using the transfer matrix method based on the model's predicted thickness parameters. Fig. 6(a) and (b) compare the predicted and measured spectra for the blue and red experimental samples, respectively. In Fig. 6(a), the predicted and measured spectra for the blue sample exhibit highly consistent linear trends in the visible range (380 nm – 760 nm), while both display low reflectance in the short-wave near-infrared region (760 nm – 1100 nm). The peaks of the predicted and measured spectra occur at wavelengths of 465 nm and 451 nm with reflectance magnitudes of 0.76 and 0.77, respectively. The inset photograph, which is a real image of the fabricated sample, confirms that the device exhibits a pure blue appearance. Likewise, Fig. 6(b) shows a strong overlap between predicted and measured spectrum for the red sample, with peaks at wavelengths of 680 nm and 662 nm and with reflectance magnitudes of approximately 0.87 and 0.82, respectively. The inset photograph reveals a pure red appearance. The high agreement between predicted and measured results is also indicated in Fig. 6(c). The predicted and measured chromaticity coordinates for the blue sample are (0.144, 0.086) and (0.149, 0.068), respectively, while for the red sample, they are (0.65, 0.34) and (0.63, 0.32). Both samples maintain very low reflectance in the short-wave near-infrared region, and their coordinates consistently lie within the high-purity regions of the chromaticity diagram. The model's predicted results align closely with experimental measurements, demonstrating its reliability and robustness in predicting high-purity color performance and minimizing short-wave near-infrared reflectance.

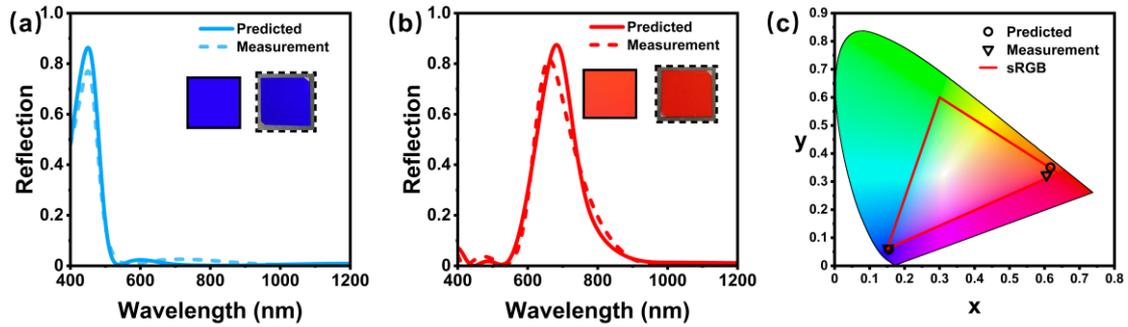

**Fig. 6** compares the reflection spectra of the experimental samples and simulation results obtained from the machine learning predictions, as well as their coordinates on the CIE-xy chromaticity diagram. The solid and dashed lines represent the predicted and experimental spectra, respectively. The inset for the solid line shows the RGB color calculated from the predicted reflectance spectrum, while the inset for the dashed line shows the image of the experimental sample. (a) The comparison of the measurement and simulation results for the blue sample. (b) The comparison of the measurement and simulation results for the red sample. (c) The chromaticity coordinates corresponding to all the reflectance spectra in (a) and (b) on the CIE-xy chromaticity diagram.

## Discussion

In this work, we demonstrate that simply increasing the depth and width of neural networks does not continuously enhance the performance of machine-learning-based inverse design models for multilayer structures. Moreover, the model exhibits large color differences when predicting high-purity colors, and it struggles to effectively control the short-wave near-infrared reflectance in the spectra. To address these issues, we develop a general optimization framework and apply three key methods in data preprocessing, model training, and inference time. First, SBC ensures uniform feature distribution in the color space. Second, ALW enhances the model's learning of high-color-difference samples. At last, PGI effectively integrates spectral information into the prediction process. Additionally, we validate the framework with different materials in the same multilayer structure, further demonstrating its generality and effectiveness. Finally, we validate the model's predictions by fabricating experimental samples and measuring their reflection spectra, which confirms their consistency with the simulation results.

Although the Color2Struct framework has already demonstrated efficiency and versatility in the inverse design of multilayer structures, we believe there remain several future research directions that can further enhance its performance and broaden its applications. These directions are summarized as follows:

(1) Model Performance Enhancement. There is still potential to boost Color2Struct by testing alternative loss-weighting strategies during training and integrating varied line-shape functions at inference time.
(2) Scaling Laws. How Color2Struct influences scaling laws during training and inference remains unclear; exploring this could reveal key insights into its behavior and efficiency.
(3) Integration with Advanced Generative Models. Incorporate Color2Struct's core ideas, including SBC, ALW, and PGI, into GANs and diffusion models to improve generation quality and convergence.
(4) Extended Structural Designs. Future work could broaden Color2Struct to more flexible design spaces, encompassing diverse material selections and their environmental responses, complex geometries beyond simple multilayer structures, and the incorporation of general fabrication or performance constraints[95–97].
(5) Applications Beyond Structural Color. SBC, ALW, and PGI are universal strategies for AI-driven science tasks with skewed output distributions, in which physical parameters are sampled uniformly but target properties vary unevenly, for example, in photonic metamaterials and topological crystals[97–99].
(6) Understanding Physics Learning. It remains unclear whether Color2Struct genuinely learns the underlying physics (such as TMM or Maxwell's equations) or merely approximates the input–output mapping; addressing this could reveal what neural inverse-design models truly capture and guide the creation of more interpretable architectures.

## Method

**Fabrication.** The metal layers (Al and Ti) are deposited by a standard e-beam evaporation method (DE Technology DE400) and the dielectric layers ($SiO_2$) are deposited by a plasma-enhanced chemical vapor deposition process (PECVD, Sentech SI 500D). The Al and Ti layers are deposited at room temperature with the rate of 2 Å/s and 0.5 Å/s, respectively, when the vacuum of the e-beam evaporator chamber was lower than $9 \times 10^{-8}$ Torr.

**Measurements.** The optical constants of all the materials were calibrated using a spectroscopic ellipsometer (RC2D, J. A. Woollam) and utilized in simulations. An ultraviolet-visible-near infrared spectrophotometer (Lambda-950) is used to measure the reflection spectra of the fabricated devices at normal incidence.

**Imaging.** A scanning electron microscope (SEM) was used to capture the cross-sectional image of the experimental sample. To obtain the cross-section, a focused ion beam (FIB) was employed to mill a groove perpendicular to the sample surface. However, ion bombardment during the milling process may cause slight alterations to the thickness of the top layers.

**Machine Learning.** All machine learning model development and training are completed using the open-source machine learning framework **_PyTorch_**.

**Computational Environment.** CPU: Intel(R) Gold 6226R; GPU: NVIDIA RTX 3090; RAM: 256 GB; Operating System: Windows 11 Pro; Python version: 3.8.

## Declarations

Availability of data and materials: Additional data can be requested from the corresponding authors upon reasonable request.

Competing Interests: The authors declare no competing financial interests.

Funding: Z. Yang acknowledges support from National Natural Science Foundation of China (12204060), and the State Key Laboratory of Information Photonics and Optical Communications (IPOC2024ZT09).
Z. Li acknowledges support from National Natural Science Foundation of China (NSFC) (No. 12404424), and Fundamental Research Funds for central universities (BUPT,2025JCTP04).
H. Ye acknowledges support from Fundamental Research Funds for the Central universities (BUPT, 2024ZCJH14).

Author contributions: Z. L., Z. Y. and J. H. conceived the project idea and supervised the project. S. H performed the machine learning simulation with the guidance from J. H.. All authors participated in the data analysis. S. H. drafted the manuscript with the help from H. Y.. All authors reviewed the manuscript.

Acknowledgements: J. Hou acknowledges inspiring discussions with C. Lu and X. Lei.

## Reference

1. Liu, Y., Zhao, T., Ju, W. & Shi, S. Materials discovery and design using machine learning. *Journal of Materiomics* **3**, 159–177 (2017).
2. Liu, R. *et al.* A predictive machine learning approach for microstructure optimization and materials design. *Sci Rep* **5**, 11551 (2015).
3. Wei, J. *et al.* Machine learning in materials science. *InfoMat* **1**, 338–358 (2019).
4. Tao, Q., Xu, P., Li, M. & Lu, W. Machine learning for perovskite materials design and discovery. *NPJ Comput Mater* **7**, 23 (2021).

5. Gubernatis, J. E. & Lookman, T. Machine learning in materials design and discovery: Examples from the present and suggestions for the future. *Phys Rev Mater* **2**, 120301 (2018).
6. Ramprasad, R., Batra, R., Pilania, G., Mannodi-Kanakkithodi, A. & Kim, C. Machine learning in materials informatics: recent applications and prospects. *NPJ Comput Mater* **3**, 54 (2017).
7. Vamathevan, J. *et al.* Applications of machine learning in drug discovery and development. *Nat Rev Drug Discov* **18**, 463–477 (2019).
8. Dara, S., Dhamercherla, S., Jadav, S. S., Babu, C. M. & Ahsan, M. J. Machine Learning in Drug Discovery: A Review. *Artif Intell Rev* **55**, 1947–1999 (2022).
9. Patel, L., Shukla, T., Huang, X., Ussery, D. W. & Wang, S. Machine Learning Methods in Drug Discovery. *Molecules* **25**, 5277 (2020).
10. Lo, Y.-C., Rensi, S. E., Torng, W. & Altman, R. B. Machine learning in chemoinformatics and drug discovery. *Drug Discov Today* **23**, 1538–1546 (2018).
11. Lo, Y.-C., Rensi, S. E., Torng, W. & Altman, R. B. Machine learning in chemoinformatics and drug discovery. *Drug Discov Today* **23**, 1538–1546 (2018).
12. Zhang, L., Tan, J., Han, D. & Zhu, H. From machine learning to deep learning: progress in machine intelligence for rational drug discovery. *Drug Discov Today* **22**, 1680–1685 (2017).
13. Ekins, S. *et al.* Exploiting machine learning for end-to-end drug discovery and development. *Nat Mater* **18**, 435–441 (2019).
14. Fang, J. A critical review of five machine learning-based algorithms for predicting protein stability changes upon mutation. *Brief Bioinform* **21**, 1285–1292 (2020).
15. Noé, F., De Fabritiis, G. & Clementi, C. Machine learning for protein folding and dynamics. *Curr Opin Struct Biol* **60**, 77–84 (2020).
16. Xie, Z.-R., Chen, J. & Wu, Y. Predicting Protein–protein Association Rates using Coarse-grained Simulation and Machine Learning. *Sci Rep* **7**, 46622 (2017).
17. AlQuraishi, M. Machine learning in protein structure prediction. *Curr Opin Chem Biol* **65**, 1–8 (2021).
18. Xu, J. Distance-based protein folding powered by deep learning. *Proceedings of the National Academy of Sciences* **116**, 16856–16865 (2019).
19. Jo, T., Hou, J., Eickholt, J. & Cheng, J. Improving Protein Fold Recognition by Deep Learning Networks. *Sci Rep* **5**, 17573 (2015).
20. Abramson, J. *et al.* Accurate structure prediction of biomolecular interactions with AlphaFold 3. *Nature* **630**, 493–500 (2024).
21. Bryant, P., Pozzati, G. & Elofsson, A. Improved prediction of protein-protein interactions using AlphaFold2. *Nat Commun* **13**, 1265 (2022).

22. Jumper, J. *et al.* Highly accurate protein structure prediction with AlphaFold. *Nature* **596**, 583–589 (2021).
23. Xu, X. *et al.* Photonic Perceptron Based on a Kerr Microcomb for High-Speed, Scalable, Optical Neural Networks. *Laser Photon Rev* **14**, (2020).
24. Mengu, D., Luo, Y., Rivenson, Y. & Ozcan, A. Analysis of Diffractive Optical Neural Networks and Their Integration With Electronic Neural Networks. *IEEE Journal of Selected Topics in Quantum Electronics* **26**, 1–14 (2020).
25. Hamerly, R., Bernstein, L., Sludds, A., Soljačić, M. & Englund, D. Large-Scale Optical Neural Networks Based on Photoelectric Multiplication. *Phys Rev X* **9**, 021032 (2019).
26. Fu, T. *et al.* Optical neural networks: progress and challenges. *Light Sci Appl* **13**, 263 (2024).
27. Williamson, I. A. D. *et al.* Reprogrammable Electro-Optic Nonlinear Activation Functions for Optical Neural Networks. *IEEE Journal of Selected Topics in Quantum Electronics* **26**, 1–12 (2020).
28. Zuo, Y. *et al.* All-optical neural network with nonlinear activation functions. *Optica* **6**, 1132 (2019).
29. Lin, X. *et al.* All-optical machine learning using diffractive deep neural networks. *Science (1979)* **361**, 1004–1008 (2018).
30. Wiecha, P. R., Arbouet, A., Girard, C. & Muskens, O. L. Deep learning in nano-photonics: inverse design and beyond. *Photonics Res* **9**, B182 (2021).
31. Wang, H. & Guo, L. J. NEUTRON: Neural particle swarm optimization for material-aware inverse design of structural color. *iScience* **25**, 104339 (2022).
32. Liu, C., Maier, S. A. & Li, G. Genetic-Algorithm-Aided Meta-Atom Multiplication for Improved Absorption and Coloration in Nanophotonics. *ACS Photonics* **7**, 1716–1722 (2020).
33. Ren, S. *et al.* Inverse deep learning methods and benchmarks for artificial electromagnetic material design. *Nanoscale* **14**, 3958–3969 (2022).
34. Ma, T., Tobah, M., Wang, H. & Guo, L. J. Benchmarking deep learning-based models on nanophotonic inverse design problems. *Opto-Electronic Science* **1**, 210012–210012 (2022).
35. Han, X., Fan, Z., Liu, Z., Li, C. & Guo, L. J. Inverse design of metasurface optical filters using deep neural network with high degrees of freedom. *InfoMat* **3**, 432–442 (2021).
36. Huang, Z., Liu, X. & Zang, J. The inverse design of structural color using machine learning. *Nanoscale* **11**, 21748–21758 (2019).
37. Ma, L., Hu, K., Wang, C., Yang, J.-Y. & Liu, L. Prediction and Inverse Design of Structural Colors of Nanoparticle Systems via Deep Neural Network. *Nanomaterials* **11**, 3339 (2021).

38. Fu, R., Chen, K., Li, Z., Yu, S. & Zheng, G. Metasurface-based nanoprinting: principle, design and advances. *Opto-Electronic Science* **1**, 220011–220011 (2022).
39. Ji, C. *et al.* Engineering Light at the Nanoscale: Structural Color Filters and Broadband Perfect Absorbers. *Adv Opt Mater* **5**, (2017).
40. Yang, B., Cheng, H., Chen, S. & Tian, J. Structural colors in metasurfaces: principle, design and applications. *Mater Chem Front* **3**, 750–761 (2019).
41. Xu, T. *et al.* Structural Colors: From Plasmonic to Carbon Nanostructures. *Small* **7**, 3128–3136 (2011).
42. Wang, D. *et al.* Structural color generation: from layered thin films to optical metasurfaces. *Nanophotonics* **12**, 1019–1081 (2023).
43. Fu, Y., Tippets, C. A., Donev, E. U. & Lopez, R. Structural colors: from natural to artificial systems. *WIREs Nanomedicine and Nanobiotechnology* **8**, 758–775 (2016).
44. Xuan, Z. *et al.* Artificial Structural Colors and Applications. *The Innovation* **2**, 100081 (2021).
45. Kinoshita, S. & Yoshioka, S. Structural Colors in Nature: The Role of Regularity and Irregularity in the Structure. *ChemPhysChem* **6**, 1442–1459 (2005).
46. Kinoshita, S., Yoshioka, S. & Miyazaki, J. Physics of structural colors. *Reports on Progress in Physics* **71**, 076401 (2008).
47. Chang, S., Guo, X. & Ni, X. Optical Metasurfaces: Progress and Applications. *Annu Rev Mater Res* **48**, 279–302 (2018).
48. Shao, L., Zhuo, X. & Wang, J. Advanced Plasmonic Materials for Dynamic Color Display. *Advanced Materials* **30**, (2018).
49. Wang, Y. *et al.* Stepwise-Nanocavity-Assisted Transmissive Color Filter Array Microprints. *Research* **2018**, (2018).
50. Zhao, Y. *et al.* Artificial Structural Color Pixels: A Review. *Materials* **10**, 944 (2017).
51. Zhao, Y., Xie, Z., Gu, H., Zhu, C. & Gu, Z. Bio-inspired variable structural color materials. *Chem Soc Rev* **41**, 3297 (2012).
52. Zhang, Z., Chen, Z., Shang, L. & Zhao, Y. Structural Color Materials from Natural Polymers. *Adv Mater Technol* **6**, (2021).
53. Flauraud, V., Reyes, M., Paniagua-Domínguez, R., Kuznetsov, A. I. & Brugger, J. Silicon Nanostructures for Bright Field Full Color Prints. *ACS Photonics* **4**, 1913–1919 (2017).
54. Tan, S. J. *et al.* Plasmonic Color Palettes for Photorealistic Printing with Aluminum Nanostructures. *Nano Lett* **14**, 4023–4029 (2014).
55. Hail, C. U., Schnoering, G., Damak, M., Poulikakos, D. & Eghlidi, H. A Plasmonic Painter's Method of Color Mixing for a Continuous Red–Green–Blue Palette. *ACS Nano* **14**, 1783–1791 (2020).
56. Zang, X. *et al.* Polarization Encoded Color Image Embedded in a Dielectric Metasurface. *Advanced Materials* **30**, (2018).

57. Ko, J. H., Yoo, Y. J., Kim, Y. J., Lee, S. & Song, Y. M. Flexible, Large-Area Covert Polarization Display Based on Ultrathin Lossy Nanocolumns on a Metal Film. *Adv Funct Mater* **30**, (2020).
58. Song, M. *et al.* Color display and encryption with a plasmonic polarizing metamirror. *Nanophotonics* **7**, 323–331 (2018).
59. Dai, P. *et al.* Inverse design of structural color: finding multiple solutions *via* conditional generative adversarial networks. *Nanophotonics* **11**, 3057–3069 (2022).
60. Dai, P., Sun, K., Muskens, O. L., de Groot, C. H. & Huang, R. Inverse design of a vanadium dioxide based dynamic structural color via conditional generative adversarial networks. *Opt Mater Express* **12**, 3970 (2022).
61. Dai, P. *et al.* Accurate inverse design of Fabry–Perot-cavity-based color filters far beyond sRGB via a bidirectional artificial neural network. *Photonics Res* **9**, B236 (2021).
62. Abd Elaziz, M. *et al.* Advanced metaheuristic optimization techniques in applications of deep neural networks: a review. *Neural Comput Appl* **33**, 14079–14099 (2021).
63. Tsakyridis, A. *et al.* Photonic neural networks and optics-informed deep learning fundamentals. *APL Photonics* **9**, (2024).
64. Hegde, R. S. Deep learning: a new tool for photonic nanostructure design. *Nanoscale Adv* **2**, 1007–1023 (2020).
65. Heidari, A., Navimipour, N. J. & Unal, M. Applications of ML/DL in the management of smart cities and societies based on new trends in information technologies: A systematic literature review. *Sustain Cities Soc* **85**, 104089 (2022).
66. Xu, S. *et al.* Deep-learning-powered photonic analog-to-digital conversion. *Light Sci Appl* **8**, 66 (2019).
67. Kaveh, M. & Mesgari, M. S. Application of Meta-Heuristic Algorithms for Training Neural Networks and Deep Learning Architectures: A Comprehensive Review. *Neural Process Lett* **55**, 4519–4622 (2023).
68. Hadi, M. U. *et al.* Large Language Models: A Comprehensive Survey of its Applications, Challenges, Limitations, and Future Prospects. Preprint at https://doi.org/10.36227/techrxiv.23589741.v6 (2024).
69. Meyer, J. G. *et al.* ChatGPT and large language models in academia: opportunities and challenges. *BioData Min* **16**, 20 (2023).
70. Chang, Y. *et al.* A Survey on Evaluation of Large Language Models. *ACM Trans Intell Syst Technol* **15**, 1–45 (2024).
71. Chang, Y. *et al.* A Survey on Evaluation of Large Language Models. *ACM Trans Intell Syst Technol* **15**, 1–45 (2024).
72. Chen, M. *et al.* Evaluating Large Language Models Trained on Code. (2021).
73. Zhao, W. X. *et al.* A Survey of Large Language Models. (2023).
74. Wei, J. *et al.* Emergent Abilities of Large Language Models. (2022).

75. Ganesan, P., Rajini, V. & Rajkumar, R. I. Segmentation and edge detection of color images using CIELAB color space and edge detectors. in *INTERACT-2010* 393–397 (IEEE, 2010). doi:10.1109/INTERACT.2010.5706186.
76. Connolly, C. & Fleiss, T. A study of efficiency and accuracy in the transformation from RGB to CIELAB color space. *IEEE Transactions on Image Processing* **6**, 1046–1048 (1997).
77. Weatherall, I. L. & Coombs, B. D. Skin Color Measurements in Terms of CIELAB Color Space Values. *Journal of Investigative Dermatology* **99**, 468–473 (1992).
78. Son, D.-K., Cho, E.-B., Moon, I.-K., Park, Y.-S. & Lee, C.-G. Development of an Illumination Measurement Device for Color Distribution Based on a CIE 1931 XYZ Sensor. *J Opt Soc Korea* **15**, 44–51 (2011).
79. Pope, R., Douglas, S., Chowdhery, A. & others. Efficiently scaling transformer inference. *Proceedings of Machine Learning and Systems* **5**, 606–624 (2023).
80. Fallah, H. R., Ghasemi, M., Hassanzadeh, A. & Steki, H. The effect of annealing on structural, electrical and optical properties of nanostructured ITO films prepared by e-beam evaporation. *Mater Res Bull* **42**, 487–496 (2007).
81. Meyyappan, M., Delzeit, L., Cassell, A. & Hash, D. Carbon nanotube growth by PECVD: a review. *Plasma Sources Sci Technol* **12**, 205–216 (2003).
82. Kamentsky, L. A., Melamed, M. R. & Derman, H. Spectrophotometer: New Instrument for Ultrarapid Cell Analysis. *Science (1979)* **150**, 630–631 (1965).
83. Vernon-Parry, K. D. Scanning electron microscopy: an introduction. *III-Vs Review* **13**, 40–44 (2000).
84. Katsidis, C. C. & Siapkas, D. I. General transfer-matrix method for optical multilayer systems with coherent, partially coherent, and incoherent interference. *Appl Opt* **41**, 3978 (2002).
85. Pettersson, L. A. A., Roman, L. S. & Inganäs, O. Modeling photocurrent action spectra of photovoltaic devices based on organic thin films. *J Appl Phys* **86**, 487–496 (1999).
86. Luce, A., Mahdavi, A., Marquardt, F. & Wankerl, H. TMM-Fast, a transfer matrix computation package for multilayer thin-film optimization: tutorial. *Journal of the Optical Society of America A* **39**, 1007 (2022).
87. Zhu, R., Luo, Z., Chen, H., Dong, Y. & Wu, S.-T. Realizing Rec 2020 color gamut with quantum dot displays. *Opt Express* **23**, 23680 (2015).
88. Guo, W. *et al.* The Impact of Luminous Properties of Red, Green, and Blue Mini-LEDs on the Color Gamut. *IEEE Trans Electron Devices* **66**, 2263–2268 (2019).

89. Broadbent, A. D. A critical review of the development of the CIE1931 RGB color-matching functions. *Color Res Appl* **29**, 267–272 (2004).
90. Habekost, M. Which color differencing equation should be used. *International Circular of Graphic Education and Research* **6**, 20–33 (2013).
91. Mokrzycki, W. S. & Tatol, M. Colour difference ΔE – A survey. *Machine Graphics and Vision* **20**, 383–411 (2011).
92. Bucher, M., Herbin, S. & Jurie, F. Hard Negative Mining for Metric Learning Based Zero-Shot Classification. in *Computer Vision – ECCV 2016 Workshops: Amsterdam, The Netherlands, October 8-10 and 15-16, 2016, Proceedings, Part III* 524–531 (Springer International Publishing, 2016).
93. Lin, T.-Y., Goyal, P., Girshick, R. B., He, K. & Dollár, P. Focal Loss for Dense Object Detection. in *Proceedings of the IEEE Conference on Computer Vision and Pattern Recognition (CVPR)* 2980–2988 (2017).
94. Xia, J., Wu, L., Wang, G. & others. ProgCL: Rethinking hard negative mining in graph contrastive learning. *arXiv preprint arXiv:2110.02027* (2021).
95. Zhang, Z. *et al.* Diffusion probabilistic model based accurate and high-degree-of-freedom metasurface inverse design. *Nanophotonics* **12**, 3871–3881 (2023).
96. Dai, P. *et al.* Inverse design of structural color: finding multiple solutions *via* conditional generative adversarial networks. *Nanophotonics* **11**, 3057–3069 (2022).
97. Wiecha, P. R., Arbouet, A., Girard, C. & Muskens, O. L. Deep learning in nano-photonics: inverse design and beyond. *Photonics Res* **9**, B182 (2021).
98. Ma, W. *et al.* Deep learning for the design of photonic structures. *Nat Photonics* **15**, 77–90 (2021).
99. Adibnia, E., Ghadrdan, M. & Mansouri-Birjandi, M. A. Nanophotonic structure inverse design for switching application using deep learning. *Sci Rep* **14**, 21094 (2024).